\documentclass[IOP]{emulateapj}

\usepackage{natbib}
\usepackage{subfigure}
\usepackage{verbatim}
\usepackage{tablefootnote}
\usepackage{graphicx}

\shorttitle{HD\,209458b: No Si\,III and the need for COS transit observations}
\shortauthors{Ballester \& Ben-Jaffel}

\begin{document}

%% LaTeX will automatically break titles if they run longer than one line. However, you may use \\ to force a line break if you desire.

\title{Re-visit of HST FUV observations of hot-Jupiter system HD\,209458:\\ 
No Si\,III detection and the need for COS transit observations}


\author{G. E. Ballester}
\affil{University of Arizona, Dept. of Planetary Sciences, Lunar \& Planetary Laboratory, \\
	 1541 E University Blvd.,  Tucson, AZ 85721-0063 USA}
    \email{gilda@lpl.arizona.edu}
    
\and
\author{L. Ben-Jaffel\altaffilmark{1}}
\affil{UPMC Univ Paris 06, UMR7095, Institut d'Astrophysique de Paris,
              F-75014 Paris, France}
\email{bjaffel@iap.fr}

\altaffiltext{1}{Visiting Research Scientist, University of Arizona, Lunar \& Planetary Laboratory.}

%% Mark off your abstract in the ``abstract'' environment. In the manuscript style, abstract will output a Received/Accepted 
%% line after the title and affiliation information. No date will appear since the author does not have this information. 
%% The dates will be filled in by the editorial office after submission.

\begin{abstract} 
 
The discovery of O\,I atoms and C\,II ions in the upper atmosphere of HD\,209458b, made with the Hubble Space Telescope Imaging Spectrograph (STIS) using the G140L grating, showed that these heavy species fill an area comparable to the planet's Roche lobe. The derived $\sim$10\% transit absorption depths require super-thermal processes and/or supersolar abundances. From subsequent Cosmic Origins Spectrograph (COS) observations, C\,II absorption was reported with tentative velocity signatures, and absorption by Si\,III ions was also claimed in disagreement with a negative STIS G140L detection.  Here, we revisit the COS dataset showing a severe limitation in the published results from having contrasted the in-transit spectrum against a stellar spectrum averaged from separate observations, at planetary phases 0.27, 0.72, and 0.49.  We find variable stellar Si\,III and C\,II emissions that were significantly depressed not only during transit but also at phase 0.27 compared to phases 0.72 and 0.49.  Their respective off-transit 7.5 and 3.1\% flux variations are large compared to their reported 8.2$\pm$1.4\% and 7.8$\pm$1.3\% transit absorptions.  Significant variations also appear in the stellar line shapes, questioning reported velocity signatures.  We furthermore present archive STIS G140M transit data consistent with no Si\,III absorption, with a negative result of 1.7$\pm$18.7 including  $\sim$15\% variability.  Silicon may still be present at lower ionization states, in parallel with the recent detection of extended magnesium, as Mg\,I atoms.   In this frame, the firm detection of O\,I and C \,II implying solar or supersolar abundances contradicts the recent inference of potential $\times$20--125 subsolar metallicity for HD 209458b. 		

\end{abstract}


\section{Introduction}

UV detection of both hydrogen and heavy species on the extended  atmospheres of hot exoplanets orbiting inside 0.1\,AU from their  stars provides a useful tool for understanding the properties of these uppermost planetary layers, the energy input from the immense stellar X-ray and UV (XUV) radiation and the processes of interaction with the incident stellar wind magnetized plasma. 
Provided information on the velocity distribution can be estimated in the transmission signature, the relative abundance of heavy species can be constrained  to in turn reveal compositional properties that also pertain to the lower atmosphere and nature of the exoplanet.  
UV studies of exoplanets can thus provide unique information for their characterization  that is usually unaccessible with IR and optical observations.
Most of what we understand about hot-Jupiter upper atmospheres is based on  far-ultraviolet (FUV)  {\em Hubble Space Telescope (HST)} transit observations and extensive modeling of  HD\,209458b, which orbits  a G0V star at 0.047\,AU.  However, as we show in this paper, the full  {\em HST} FUV observational potential of this target has yet to be reached. Using the Space Telescope Imaging Spectrograph (STIS) with the G140M medium-resolution (R $\sim$ 10,000) grating, a $\sim$10\% absorption in Lyman-$\alpha$ was discovered by H\,I  atoms in an inflated atmosphere reaching the Roche lobe distance of $\sim 2.9$\,$R_{pl}$ \citep{vid03,vid08,ben07,ben08}.  	
A Lyman-$\alpha$ absorption with an enhanced blue-shifted component was first reported from the average of three medium-resolution G140M transits by \citet{vid03}, while \citet{ben08},  who included a fourth partial transit, found a more symmetric average absorption in the red and  blue wings. Existing theoretical models have tried to reproduce the  H\,I observations assuming the solar XUV input where the atmospheric inflation results from heating at the base of the thermosphere around 0.01-1\,$\mu$bar,  yielding a mean temperature of order 10,000\,K that drives a hydrodynamic outflow. 
Most models find enough thermal atomic hydrogen  to reproduce the symmetric absorption profile originally reported in \citet{ben08}  for the HD\,209458b transits \citep{ben08,ben10,tra11,kos10,kos13a}.   However, some issues remain unresolved regarding the exact total thermal H content that depends on the atmospheric solar or supersolar abundances assumed \citep{ben10,kos10,kos13a,kos13b}, 	on the strength of the outflow and the tidal effects \citep[e.g.,][]{lam03,yel04,tia05,gar07,mur09,sto09,guo11,guo13,bou14}, on the geometrical distribution of the gas which can be global or quite likely confined at low (magnetic) latitudes due to the presence of a planetary magnetic field  \citep{tra11,tra14,ada11,kho12,coh12,owe14}, as well as on the presence of non-thermal processes affecting the energetics and line broadening  \citep{ben10},and on the strength of the stellar XUV heating \citep{bal15}.  Charge exchange with the incident stellar wind protons  \citep{hol08,tre13,boulec13} and stellar radiation pressure \citep{vid03}  are also key mechanisms that predict an asymmetric Lyman-$\alpha$ transit absorption profile, but the relevance of these processes depends on the inclusion of natural line broadening \citep{ben08}	and self-shielding \citep{boulec13}, and a  comprehensive analysis is required including the presence of a planetary magnetic field \citep{kis14} and the inherent absorption by the interstellar medium thay may erase the asymmetric absorption feature if the Doppler shift is not strong enough.
Probably the interstellar medium (ISM) absorption and the relatively low signal-to-noise ratio (S/N) limit the H\,I Lyman-$\alpha$ data analysis and interpretation, yet all studies agree on the important result that an area equivalent to the size of the Roche lobe is filled on HD\,209458b, a finding that has many implications on our understanding of these atmospheres when connected with transit signatures in other species 
\citep{ben10}.	 
However, we must stress here that available observations already show that we are close to making a final diagnostic on the very nature of the transit absorption as shown by the comparison between light curves derived for the blue and red wings of the H\,I Lyman-$\alpha$ line \citep[e.g., Fig. 3 in][]{ben08}.  For the relatively good signal to noise reported thus far  \citep{ben08}, repeated new STIS G140M medium-resolution transit observations of HD\,209458b should help converge our determination of the H\,I Lyman-$\alpha$ transit absorption profile for this planet, and definitely elucidate any asymmetric spectral absorption signature.  Additionally,  time variation may be present in the H\,I Lyman-$\alpha$ spectral signature of HD\,209458b, as can be seen in Fig. 3b of \citep{ben07}.  Similarly, time variation  has been clearly identified in the H\,I Lyman-$\alpha$ absorption of the hot-Jupiter HD\,189733b \citep{lec12,bou13}.   On HD\,189733b, one STIS observation  showed no H\,I Lyman-$\alpha$ absorption while on another date it showed absorption that was   enhanced in the blue wing.   A STIS observation of  GJ\,436b has also shown an extended hydrogen atmosphere on this warm Neptune, revealing a post-egress signature from a hydrogen tail with an enhanced blue-shifted H\,I Lyman-$\alpha$ absorption  \citep{kul14}.

On HD\,209458b, the STIS detection of extended neutral oxygen atoms and singly ionized carbon (the two dominant forms of the species for this environment)  also  extending to near  Roche lobe distances has also been of great importance,  since these components can provide additional critical  diagnostics of the upper atmosphere \citep{vid04,ben10}.  Their detection shows that  heavy species are effectively entrained in the hydrodynamic outflow of the lighter  hydrogen gas  (through collisions with the neutral hydrogen atoms and the protons) and overcome  diffusive separation  \citep{gar07,kos10,kos13a,kos13b}.  The transit absorption depths in the O\,I 1304\,\AA\ and C\,II 1335\,\AA\  multiplets derived from the STIS G140L observations of HD\,209458b were,  respectively, $12.8\pm4.5$\% and $7.5\pm3.5$\%  \citep{vid04}, or $10.5\pm4.4$\% and $7.4\pm4.7$\% from a re-visit of the observations \citep{ben10} and correspond to  2.8 and 2.2-sigma, or 2.4 and 1.6-sigma detections. These large absorption depths cannot be explained  by 10,000\,K thermal populations at solar abundances since  line profiles that would result from thermal and natural broadening for vertical distributions from 1-D hydrodynamic models  are not wide enough to significantly absorb the broad stellar lines (with solar line widths) during transit.  This is explored in various papers (see below).  Since the  line absorption  profiles and thus the velocity distributions of the absorbers were unresolved  with STIS (using the G140L grating at R$\sim$1000),  substantially different interpretations have been put forth. 

To broaden the absorption profile while keeping solar abundances, energetic super-thermal  populations have been invoked (with effective temperatures $\times$5--100 larger than a $\sim$10,000\,K background H-dominated gas), such as  from processes that preferentially energize minor species involving chemistry, radiation, and wave-particle interactions,  or from turbulence and the stellar wind interaction \citep{ben10}.  Ben-Jaffel \& Hosseini have shown that the final transit absorption profile is strongly dependent on the vertical distribution of the super-thermal population, a diagnostic that could be used if medium-resolution transit spectra are obtained.  Alternatively, to enhance the optical thickness of the thermal population, some hydrodynamic models find that $\times$3--5 supersolar abundances would be required accompanied by a hotter thermosphere from twice the stellar XUV input than from the mean Sun \citep{kos13b}.   
\citep[A previous hydrostatic approximation invoked quite large {$\times$4--40} supersolar abundances with an unreasonably strong XUV flux  {$\times$5--10 } higher than the mean Sun;][]{kos10}.   However, recent findings  are now showing an unexpected long-term low activity on the star HD\,209458, and this low activity does not support scenarios that  invoke even moderately high XUV fluxes, such as those of solar mean or maximum activity, to heat and inflate the atmosphere  to the extent required by the observations \citep{bal15}. As there  appears to be an energy crisis in the upper atmosphere of this planet  due to  the low XUV input, new information on super-thermal processes that may operate in the upper atmosphere of this planet are needed, and such evidence is uniquely accessible with  transit observations that resolve the line absorption profiles and thus velocity and energy distribution of the heavy species.
 
The Cosmic Origins Spectrograph (COS) on {\em HST} provides higher sensitivity than STIS, and it can improve the S/N of the transit absorption in C\,II 1335\,\AA\   and in O\,I 1304\,\AA\ as well as resolve the multiplets and the line absorption profiles at medium spectral resolution (with the G130M grating at R$\sim$20,000) as has been demonstrated for the case of HD\,189733b \citep{ben13,ben14}.  FUV observations of the HD\,209458 system were made with  COS G130M by  \citet{lin10}. Absorption by the planet during transit was reported in the C\,II 1335\,\AA\ doublet lines  with a depth of  $7.8\pm1.3$\% in the  --50 to +50\,km\,s$^{-1}$ interval, similar  to the line-integrated absorption detected with STIS.  Tentative velocity signatures were  reported,  with  absorption at --10 to 15\,km\,s$^{-1}$ and less certain features at --40 and +30--70\,km\,s$^{-1}$, and no absorption at zero velocity. The larger velocities would agree with the presence of an upper atmospheric hot population, or of  plasma  with large motions around the planet \citep{lin10,kos13b}, with great  implications about  the properties of HD\,209458b's upper atmosphere.  

From the same COS dataset, \citet{lin10} also  reported an $8.2\pm1.4$\% transit absorption by Si\,III ions in the 1206.5\,\AA\ line.  In sharp contrast, no Si\,III 1206.5\,\AA\ absorption was detected in the average of the four {\em HST} transit observations made with STIS G140L that detected O\,I and C\,II, where a negative result of  0.0$^{+2.2}_{-0.0}$\%  was derived in Si\,III  \citep{vid04}.  To fit the COS Si\,III results,  \citet{kos13b}  found similar  $\times$3--5 supersolar abundances and higher stellar energy input requirements for the upper atmosphere, similar to the inferences made for the O\,I and C\,II.  

The composition at high atmospheric altitudes (as modified by the stellar ionization, thermal balance, and charge-exchange processes)  reflects the species and abundances at the base of the thermosphere, and it is thus diagnostic of the properties of the  lower atmosphere and overall planetary abundances and energy budget  \citep[e.g.,][]{gar07,kos13a,kos13b,lav14}. This is particularly the case for a hot Jupiter with its strong  3D atmospheric circulation and large effective eddy mixing  \citep[e.g.,][]{sho09,par13}, and where the detection of extended O\,I and C\,II on HD\,209458b itself shows that diffusive separation of the heavy species in the upper atmosphere is impeded by the hydrodynamic outflow  \citep{kos13a,kos13b}.  For O\,I and C\,II the key lower atmospheric components  are H$_2$O,  that does not condense on hot Jupiters, and CO,  the dominant carbon-bearing species on hot Jupiters \citep[e.g.,][]{mos11}.  For refractory  species like silicon, the situation is more complex and uncertain. For silicon to be present in the thermosphere, \citet{kos13b} find that it cannot condense into silicate clouds  in the  lower atmosphere, and that the vertical transport should be strong to keep any condensate particles lofted. Furthermore, silicon must be predominantly in the Si\,II ionization state, and charge exchange with protons (and He$^+$) can then supply a significant Si\,III population to be detected. It may well be that silicon is not effectively  removed by vertical and horizontal cold-traps on this planet, given the recent detection of Mg\,I in the upper atmosphere \citep{vid13}, but we must  first address the validity of the reported COS detection of silicon ions, at least as  Si\,III, given the contradiction with the negative STIS G140L results.  

In this paper we re-visit the COS observations of the HD\,209458 system given that these observations were not made in the standard fashion of observing immediately before and/or after the transit to contrast the in-transit fluxes with the actual stellar fluxes around the time of transit.  Instead, the in-transit stellar fluxes were ratioed against the average fluxes measured on three separate HST visits that were obtained over a period of one to two weeks with the planet off transit. The limitation of this approach is that the stellar activity, even if it is low, can cause significant flux variations in the FUV line emissions such as over a stellar  rotation period, and from stochastic or flare activity \citep[e.g.,][]{par14}.   This is indeed the case for the Sun for which a variation of $\sim 27$\% was reported for the Si III 1206.5\,\AA\ line during a rotation period \citep[e.g., Table 3 in][]{ben13}.   It is thus possible that the transit absorptions derived from the COS data could be related to stellar variation rather than to the true planetary transit signature.    \citet{lin10} deduced their approach  to be valid because the flux level and line shape in the Si\,IV 1393.7\,\AA\ line was similar in the in-transit data and in the averaged off-transit data.  We find, however, that this  use of the averaged off-transit data requires reassessment.  

In Sec. 2 we  re-evaluate the COS observations in the Si\,III 1206.5\,\AA\ line, C\,II 1335\,\AA\ doublet,  Si\,IV 1393.7\,\AA\ line, and 1364.8--1390.7 and 1405.6--1423.6\,\AA\ continua,  taking advantage of the temporal information of the data, both from the time-tagging and from comparison of the average stellar fluxes per  HST visit.   We find indeed significant time variation in the stellar emissions that render the previously derived transit absorptions for Si\,III 1206.5\,\AA\ line and the C\,II 1335\,\AA\ doublet from this COS dataset unreliable.
Complications in the usage of the COS dataset for a derivation of velocity signatures in the absorption are also addressed. Moreover, we also present previously unpublished Si\,III transit observations of HD\,209458b from a re-visit of the original STIS G140M dataset that discovered  extended H\,I in Lyman-$\alpha$ absorption, since these exposures also sampled the Si\,III 1206.5\,\AA\ line. We find no evidence of a Si\,III transit absorption, in agreement with the  first STIS G140L findings.  We present and discuss the STIS data in Sec. 3. We also discuss some issues related to the conditions that would be needed in the lower atmosphere to obtain silicon ions in the upper atmosphere.  We finish by establishing the need for COS FUV transit observations of this exoplanet, in particular for the unexploited diagnostics in the O\,I 1304\,\AA\ and C\,II 1335\,\AA\ transit signatures.   

\section{COS observations}

\subsection{Observations and data reduction}

HD\,209458 was observed with COS on Sep--Oct 2009 during four HST visits,  two or three HST orbits each, with the planet around orbital phases $\phi$=0.27, 0.72, 0.0, and 0.49 (Table 1). The COS G130M grating was used covering the $\sim$1130--1460\,\AA\ region at medium spectral resolution (R$\sim$20,000 $\sim$15\,km/s, with 0.01\,\AA\ sampling). Throughout each visit the exposures were made  alternating between four grating central wavelength settings referred to as 1291, 1300, 1309, and 1318. COS FUV spectra are sampled by two detector segments, A and B, that are separated by a small gap. Some of the grating settings either did not sample the important O\,I\,1304\,\AA\ triplet or had the lines near the edge of the detector (Table 1), so this stellar emission could not be evaluated as was possible for transit observations of HD\,189733b \citep{ben13,ben14}.   Note also that the C\,II 1335\,\AA\ feature is a triplet, but it is detected as a doublet with a  line at 1334.5\,\AA\ and  two unresolved lines at 1335.7\,\AA. 

%Table 1

\begin{table}\label{tab:obs}
\caption[]{Summary of COS observations}
\centering
\begin{tabular}{c c c c c c }
\hline\hline
 Exp. &Start  time   & Exp.     & Planet	& Grating \\		
 No.   & in 2009     & time     & mid.	&  wavelength \\		
 ~       & (UT)           & (sec)    & orbital	& setting$^a$  \\	
 ~       &                    &              &  phase     \\
 \hline 
 01 & Sep 19  10:10 & 2340 & 0.255 & 1291 \\	
 02 & Sep 19  11:36 & 0955 & 0.270 & 1300 \\	
 03 & Sep 19  11:55 & 1851 & 0.275 & 1309\\	
 04 & Sep 19  13:11 & 0560 & 0.288 & 1300\\	
 05 & Sep 19  13:24 & 2235 & 0.294 & 1318\\	
\\
 06 & Sep 24  13:12 & 2340 & 0.710 & 1291\\	
 07 & Sep 24  14:38 & 0945 & 0.724 & 1300\\	
 08 & Sep 24  14:57 & 1787 & 0.730 & 1309\\	
 09 & Sep 24  16:14 & 0925 & 0.743 & 1300\\	
 10 & Sep 24  16:33 & 1798 & 0.748 & 1319\\	
\\
 11 & Oct 02  14:32 & 1045 & 0.993 & 1291\\	
 12 & Oct 02  14:53 & 1096 & 0.997 & 1300\\	
 13 & Oct 02  15:55 & 1400 & 0.010 & 1309\\	
 14 & Oct 02  16:22 & 1405 & 0.015 & 1319\\	
\\
 15 & Oct 18  10:51 & 1057 & 0.490 & 1291\\	
 16 & Oct 18  11:18 & 1093 & 0.494 & 1300\\	
 17 & Oct 18  12:20 & 1401 & 0.507 & 1309\\	
 18 & Oct 18  12:46 & 1405 & 0.512 & 1319\\	
 \hline \hline
\end{tabular}
\begin{list}{}{}
   \item[$^{a}$] 
The spectral ranges covered per grating wavelength setting respectively for detector Segments B \& A are: 
(1291) 1132-1274 \& 1291--1433\,\AA;
(1300) 1141-1283 \&1300-1442\,\AA;
(1309) 1153-1294 \& 1309-1449\,\AA; and 
(1318) 1163-1303 \&1319-1459\,\AA.
\end{list}
\end{table}

The  time-tagged exposures were  reprocessed  into 100\,s segments, but since the  S/N was  low, the segments were averaged into 1000\,s  sub-exposures so that the temporal flux variations (depicted in Fig.~1) were not dominated by the photon noise level. For the extraction of the  stellar  emission line fluxes, a weak stellar continuum was  first subtracted from each spectrum based on a linear fit to the spectral regions devoid of obvious emission lines identified in a high-resolution solar spectrum \citep{cur01}.  For the Si\,III\,1206.5\,\AA\ line  we also subtracted a linear fit of the blue wing of the stellar H\,I Lyman-$\alpha$ line which is very broad as detected by COS.  
This is a relatively small correction compared to the potential planetary absorption, and the associated errors have been included in the propagated errors. 
 The stellar line fluxes,  depicted in Fig. 1, were  extracted  by co-aligning the given line in all the spectra, and  integrating the spectral fluxes out to about 10\% of the averaged peak intensity  \citep{ben07}.  This integration spanned about +/--28\,km/s. Photon noise errors were propagated in all reduction steps and problematic pixels (with dq\_wgt=0) were ignored.

%________________________________________________________________________
\subsection{Absolute Wavelength Scale} 

The determination of a common absolute wavelength scale, and thus of any  planetary velocity signature reported from the COS dataset, was complicated because of having alternated between the four G130M  central wavelength settings during each visit. The original purpose of that setting was to minimize fixed-pattern noise and grid-wire shadow effects \citep{fra10,lin10},  while for most other transit  observations of exoplanets keeping  instrumental settings fixed has proven best for measuring relative changes due  to the planetary transmission alone. The grid-wire shadow effects are now automatically corrected by the COS pipeline software, as in the  CALCOS 2.14.4 and 2.18.5 pipeline versions used for the time-tagged flux analysis and the wavelength study, respectively. 
The main complication is that the repositioning of the grating is not exactly repeatable due to thermal flexures \citep{sha09}.  Although in principle the absolute wavelength scale for each exposure could be set by co-aligning the C\,II 1334.5\,\AA\ line  based on the dip from the ISM  absorption of this line as done by \citet{lin10},  we found that the S/N is too low to accurately find the center of this absorption in each exposure.  To co-align the spectra, for the Segment-A data we compared co-aligning the spectra against each line of the C\,II 1335\,\AA\  doublet  and also against the full doublet. Significant differences were found when using only one line or the full doublet, when the data were smoothed or not, and when different exposures were chosen as the template.   
Comparing results from fitting against exposures 1 and 15  (Table 1), using only the C\,II 1334.5\,\AA\ line or the full doublet, and with or without smoothing, we found offsets in the centering of the ISM 1334.5\,\AA\ line absorption of about 10\,km\,s$^{-1}$ (about 5 pixels).  This error, though somewhat smaller than the 15\,km\,s$^{-1}$ spectral resolution, produces large differences in the resulting line profiles and in the corresponding transit absorption line profile. In Fig. 2 (Sec. 2.4) we show a sample result  of line profiles from  co-alignment against exposure 15, using the full doublet, and with the data box-car smoothed by 5.

The Si\,III\,1206.5\,\AA\ line data, sampled by the detector Segment B, was co-aligned separately. The wavelength shift found from the C\,II 1335\,\AA\ doublet in the Segment A data differed from that found for  Segment B by up to $\pm$4  pixels (while Linsky et al. 2010 applied the same shift to data from both segments). This turned out to be expected since the COS pipeline calibration calculates the spectral offset for each detector segment independently. The SHIFT1A and SHIFT1B in the processed data differ by about 4--7 pixels, similar to the  offsets that we found.

\subsection{Time variation of the stellar fluxes}  

Figure 1 shows the time-resolved  fluxes for the  Si\,III 1206.5\,\AA\ line, C\,II 1335\,\AA\ doublet,  Si\,IV 1393.7\,\AA\ line, and 1364.8--1390.7 and 1405.6--1423.6\,\AA\ continua.  Black symbols show the fluxes from the 1000\,s sub-exposures, and gray symbols show the averaged fluxes per HST visit. For each  feature, the fluxes have been normalized by the corresponding average value for visit 4 ($\phi\sim$0.49) to ease the interpretation of the results. The error bars for the sub-exposures are propagated statistical photon noise, while the error bars for the visit-averaged fluxes include the non-statistical flux scattering within a given visit  defined as  the standard deviation from the mean.  The variations within an HST visit can be stellar  but there can also be instrumental effects \citep[e.g.,][]{lin12}.  Since the intra-orbit flux variations do not seem repeatable from orbit to orbit, they cannot be systematically corrected for.  The stellar fluxes averaged per visit are provided in Table 2 with both the photon-noise  errors as well as  errors that include the intra-visit scattering.  In the discussion below we refer to the photon-noise errors for direct comparison with the results by \citet{lin10}.

%Figure 1

\begin{figure}\label{figLotfi}
   \centering
    \hspace*{-0.15in}
    	\includegraphics[width=3.75in] {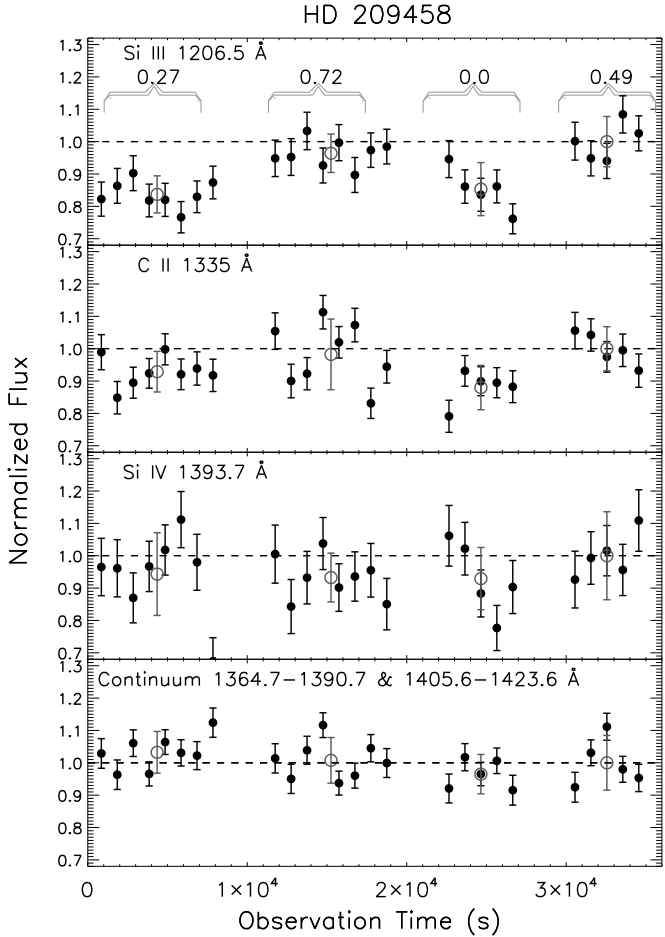} %{ballester-fig1.eps}	%
    \caption{Normalized fluxes for the 1000\,s sub-exposures versus observation time for the stellar emission features  Si\,III 1206.5\,\AA\ line,   C\,II 1335\,\AA\ doublet,  Si\,IV 1393.7\,\AA\ line, and the 1364.8--1390.7 \& 1405.6--1423.6\,\AA\ continuum. The four HST visits are labeled on the top panel by the  planetary orbital phase. For each visit  the times have been offset from their arbitrary zero-time  by  0, 1$\times10^4$, 2$\times10^4$, and 3$\times10^4$\,s, respectively. For each  emission feature, the average flux per visit is over-plotted as a gray circle at the middle of the visit.}
              \label{FigLotfi}
\end{figure}

%Table 2

\begin{table}\label{tab:fluxes}
\caption[]{Normalized stellar fluxes from COS dataset, averaged per HST visit}
\centering
\begin{tabular}{l ccccc }
\hline\hline
 Spectral   & Planet           & Norm. 	& Photon  & {Intra-} & Photon \\
 feature     & mid.           &  flux$^a$     &noise      & {visit}  & and \\
 ~               &   orb.                              &                       &error      &  {std}  &  intra-\\
 ~               & phase                               &                       &               &  {error$^b$}  &  visit\\
 ~               &                                &                       &                &   &error$^b$\\
 \hline 
{Si\,III 1206.5\,\AA}
   & $\phi=0.27$   &  0.837 &0.018   & 0.054  & 0.057  \\
line 
   & $\phi=0.72$   &  0.964 &0.020   & 0.056  & 0.059  \\
~ & $\phi=0.00$   &  0.853 &0.023   & 0.079  & 0.082  \\
~ & $\phi=0.49$   &  1.000 &0.025   & 0.074  & 0.078  \\
\hline
{C\,II 1335\,\AA}
   & $\phi=0.27$   &   0.929 &0.017  &  0.060 &  0.063  \\
doublet 
   & $\phi=0.72$   &   0.982 &0.018  &  0.108  &  0.109  \\
~ & $\phi=0.00$   &   0.880 &0.021  &  0.065  &  0.069  \\
~ & $\phi=0.49$   &   1.000 &0.023  &  0.064 &  0.068  \\
 \hline 
 {Si\,IV 1393.7\,\AA} 
   & $\phi=0.27$   &    0.943 &0.029  &  0.125  &  0.128  \\
line 
   & $\phi=0.72$   &    0.933 &0.029  &  0.070  &  0.075  \\
~ & $\phi=0.00$   &    0.929 &0.036  &  0.089  &  0.096  \\
~ & $\phi=0.49$   &    1.000 &0.038  &  0.131  &  0.136  \\
 \hline 
{1364.8--1390.7 \&} 
   & $\phi=0.27$    &   1.032 &0.015  &  0.063  &  0.064  \\
{1405.6--1432.6\,\AA}
   & $\phi=0.72$    &   1.008 &0.015  &  0.069  &  0.070  \\
continuum
   & $\phi=0.00$    &   0.965 &0.019  &  0.058  &  0.061  \\
~ & $\phi=0.49$    &   1.000 &0.019  &  0.083  &  0.085  \\
\hline\hline
\end{tabular}
\begin{list}{}{}
  \item[$^{a}$] The fluxes have been normalized with respect to the average of visit 4 ($\phi=0.49$), given by: 
 $2.12\times10^{-15}$ for  Si\,III;  $2.33\times10^{-15}$  for  C\,II;  
   $0.96\times10^{-15}$  for  Si\,IV, and;  $4.87\times10^{-15}$\,ergs\,cm$^{-2}$\,s$^{-1}$ for the continuum.
  \item[$^{b}$] The intra-visit error is the standard deviation of the  scattering of the stellar flux during the given visit.
\end{list}
\end{table}

To evaluate the validity of deriving a transit absorption from this COS dataset we look at the normalized fluxes averaged per HST visit (Table 2). For the C\,II 1335\,\AA\ doublet we find that although the flux during transit ($\phi=0.0$) is low at 0.880$\pm$0.021, the flux is also relatively low for $\phi=0.27$ at 0.929$\pm$0.017.    In contrast, the flux  for $\phi=0.72$  of 0.982$\pm$0.018 is similar to the reference value of 1.0$\pm$0.023   for $\phi=0.49$ within the errors. Although the  transit depth relative to the mean of the off-transit fluxes  (at $\phi=0.27$, 0.72, and 0.49) would be  9.3$\pm$2.4\%, which is somewhat deeper than the  $7.8\pm1.3$\% reported by \citet{lin10}, the range of depths relative to the different average fluxes for the off-transit visits  spans from 5.3$\pm$2.9\% to 12.0$\pm$2.9\%. The detection and  depth of the C\,II 1335\,\AA\ transit signature of HD\,209458b  associated thus far with the COS instrument remains unconfirmed given the inappropriate observing method that was employed.
 
The Si\,III 1206.5\,\AA\ line also shows a strong time variation. A low  flux of 0.837$\pm$0.018  is observed at $\phi=0.27$ and a similar value of 0.853$\pm$0.023 is seen during transit. Furthermore,  the flux of 0.964$\pm$0.020  observed at $\phi=0.72$  is comparable to the reference value of 1.00$\pm$0.025 at $\phi=0.49$ within the errors.  Deriving a  transit absorption for Si\,III is  therefore misleading, given the large time variations.  A detection of Si$^{++}$ ions in the upper atmosphere of  HD\,209458b is not valid from this COS dataset.

The 1364.8--1390.7 \& 1405.6--1432.6\,\AA\ continuum consists of the photospheric continuum (that decreases towards shorter wavelengths) and a white continuum emission from  the chromosphere \citep[that increases towards shorter wavelengths;][]{lin12}.  No large continuum-flux variations are observed in tandem with the line emissions, indicating that no large common-mode instrumental effect was at play. This chromospheric component is  expected to vary with stellar activity, yet the variability at these wavelengths should be smaller than in any of the emission lines \citep[e.g.,][]{sno10}.  Although in the observation at $\phi=0.27$  the continuum did not appear relatively low as in the other features but was instead somewhat larger,  and during the transit visit it was relatively low as could be due to the transit of the planetary disk, the fluxes are all similar within the errors.  

For the Si\,IV 1393.7\,\AA\ line (significantly brighter than the other 1402.8\,\AA\ line of the doublet), we find that the fluxes are relatively low not only at both $\phi=$0.0 and 0.27, but also at $\phi=0.72$ unlike for Si\,III and C\,II for which the fluxes at $\phi=0.72$ were higher and comparable to the fluxes at  $\phi=0.49$.   Thus, the stellar  Si\,IV emission turns out not to be an accurate proxy for the  C\,II and Si\,III emissions.

%Table 3

\begin{table}\label{tab:va1}
\caption[]{HD\,209458 average stellar fluxes from COS dataset, and apparent visit-to-visit variations from  off-transit data}
\centering
\begin{tabular}{l ccc }
\hline\hline
 Spectral   					& Flux$^{a,b}$			& \%Variation$^b$ &\%Variation$^b$\\	  
 feature    					& ($10^{-15}$				& as a  			& as a		\\  
    ~              					& ergs/cm$^{2}$/s)	& std. dev. 		& max/min	\\ 
% \hline \hline
 \hline 
{Si\,III 1206.5\,\AA} line			 &1.98$\pm$0.15	& 7.5\%		& 19.4$\pm$8.3\%	 	\\ 
{C\,II 1335\,\AA} doublet			 &2.26$\pm$0.07	& 3.1\%		& ~7.7$\pm$5.4\%		\\ 
{Si\,IV 1393.7\,\AA}  line 		 	 &1.47$\pm$0.05	& 3.1\%		& ~~7.2$\pm$10.7\%	\\ 
{1364.8--1390.7 \&}			 		 &4.94$\pm$0.07	& 1.4\%		& ~3.2$\pm$5.0\%	\\ 
{1405.6--1432.6\,\AA} cont.\\
\hline\hline
\end{tabular}
\begin{list}{}{}
 \item[$^{a}$] Mean spectral flux and standard deviation. 
 \item[$^{b}$]  Errors include intra-visit noise.
 \end{list}
\end{table}

The variations of the emissions as observed in the off-transit visits are listed in Table 3. Clearly,  the 7.5\% variation  (or 19.4$\pm$8.3\%  max/min) of the Si\,III 1206.5\,\AA\ flux is quite large, more so than for the C\,II 1335\,\AA\  or Si\,IV 1393.7\,\AA\  fluxes although there may be some overlap if the intra-visit scatter is included in the errors.

It is  interesting to note  that the  3.1\% variation (or 7.2$\pm$10.7\% max/min) of the  Si\,IV 1393.7\,\AA\ line is smaller than the 7.5\% variation of the  Si\,III 1206.5\,\AA\  line (or 19.4$\pm$8.3\% max/min)  for the same set of observations (although again they overlap within the larger error bars). Since the Si\,IV 1393.7\,\AA\  emission originates from a hotter layer in the stellar atmosphere (at log\,$T=4.75$ for the Sun) than the C\,II 1335\,\AA\  and Si\,III 1206.5\,\AA\  emissions  \citep[at log\,$T=4.10$  and 4.25, respectively;][]{woo00}  it would be expected in general to vary more strongly or at least comparably to the Si\,III 1206.5\,\AA\ line (based on more variability at higher temperature regions).

Solar observations indicate that  the  Si\,IV 1398\,\AA\  doublet variability  is comparable in general to that of the Si\,III 1206.5\,\AA\  line, or somewhat larger at high activity as  known from short-term  and long-term observations and from observations of flares \citep{woo00,bre96}. However,  a recent evaluation of the solar variability in the FUV emission lines relevant to hot-Jupiter transit studies has been made by   \citet{ben13} using 2003--2007 spectra from the Solar Stellar Irradiance Experiment  (SOLSTICE) instrument on-board the {\em Solar Radiation and Climate Experiment (SORCE)} obtaining more detailed or relevant results for exoplanet FUV {\em HST} work. The data were divided into  periods of low, medium, high, and extreme flare activity, where the lower activity should include the emission from plages and enhanced network.  In three out of four flare level categories, the  variability of the  Si\,IV 1398\,\AA\  doublet is comparable to that of the Si\,III 1206.5\,\AA\ line,  and somewhat larger for the cases of high activity in agreement with the findings by \citet{woo00}.   In that study, there was a separate evaluation for the Si\,IV 1393.7\,\AA\  {\em line} \citep[see bottom part of Table 3 in][]{ben13}. The solar Si\,IV 1393.7\,\AA\  line showed, in contrast, that the variability with solar 27-day  rotation and 11-year cycle was somewhat smaller, at 22 and 60\%, respectively, compared to 27 and 73\% for the Si\,III 1206.5\,\AA\ line.  (These variabilities are also standard deviations from the mean.) The solar lines do show more variability that the COS results for HD\,209458, but the COS dataset is extremely limited to three samples.  
Here we present results for the Si\,IV 1393.7\,\AA\ line and not the doublet, since this was the line  used by \citet{lin10} as a proxy for the stellar activity given that it is  about twice as bright as the second line in the doublet.   In a separate analysis of the COS dataset (not presented here) in which we have not divided the  exposures into time-tagged segments, we see that the variation in the full doublet is larger and more comparable to that of the Si\,III 1206.5\,\AA\  line, due to inclusion of the more variable and/or noisier second line in the doublet.   \citet{par14} report on temporal variations on FUV fluxes of sunlike stars in 60-second time intervals. For HD\,209458 they report  upper limits for the variation in the Si\,III 1206.5\,\AA\ line and Si\,IV 1398\AA\ doublet emissions (which is not surprising given the relatively low S/N of the data), and their upper limit variation for the Si\,IV doublet is a bit larger than for the Si\,III line.  Therefore, from the various lines of evidence presented, it is unclear at this point if the relatively low visit-to-visit variation of the Si\,IV 1393.7\,\AA\  line compared to the larger variation in the Si\,III 1206.5\,\AA\  line is significant and distinct for this star compared to the Sun.
The degree of intra-visit scatter for both features seems comparable,  however, although the COS data is extremely limited.
Future  observations should shed more light into the simultaneous temporal FUV flux variations on this star.

Finally, it is also interesting that the  Si\,III 1206.5\,\AA\ and C\,II 1335\,\AA\ emissions may have shown a stellar rotational variation. Visits 1, 2, 3, and 4 were made on days 0 (reference date), 5, 13, and 29.  Using the stellar rotation periods of $\sim11.2$ -- 14.1\,days derived from the Rossiter-McLaughlin effect and the stellar line widths provided in the literature \citep[c.f.,][]{bal15},  these visits corresponded to stellar rotational phases of 0 (reference phase), 131--165,  336--63, and 20--212$^\circ$.  Therefore, visits 1 and 3 that showed the lowest fluxes should correspond to similar stellar phases, around the reference phase 0$^\circ$.

%Figure 2

\begin{figure*}\label{fig:shapes}
   \centering
    \hspace*{-0.15in}
    		\includegraphics[width=5.5in,angle=0] {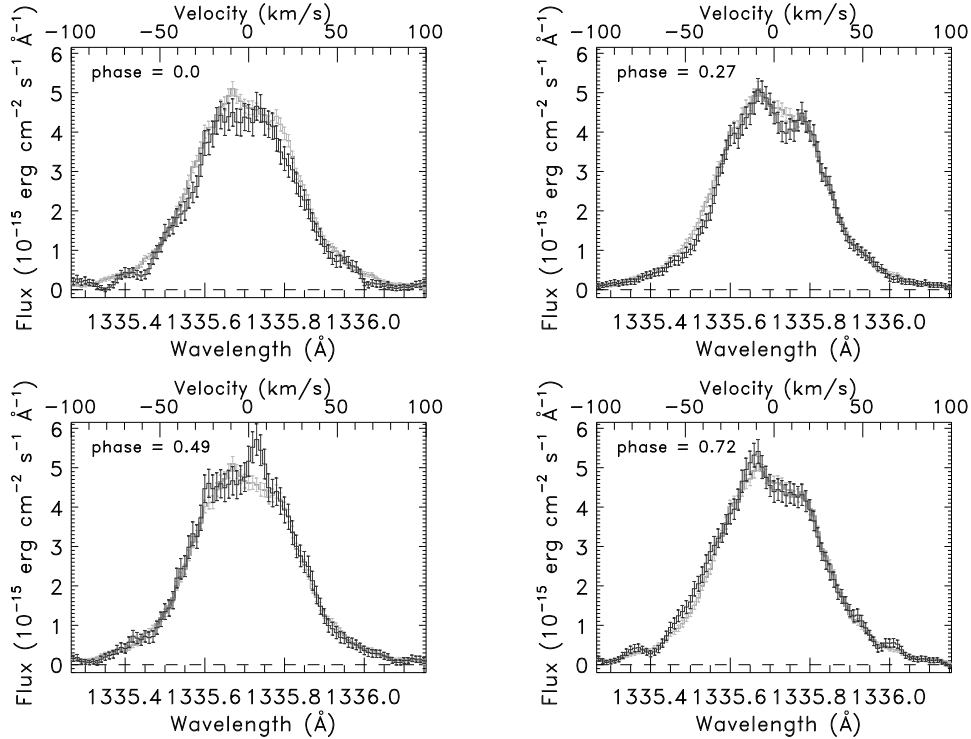} % {ballester-fig2.eps}	%
    \caption{Sample  shapes of  the  C\,II 1335.7\,\AA\ line derived from the  COS data of the four {\em HST} visits, box-car smoothed by 5, with error bars.
    The  gray   profile is the average off-transit data ($\phi=0.27$, 0.49, 0.72).  
   To correct for non-repeatable  wavelength  offsets from alternating the grating setting, 
        the data were co-aligned against exposure \#15, using both  lines of the C\,II 1335\,\AA\ doublet.  
        Visit-to-visit variation in the stellar line shape  is apparent, but observations with a single grating setting should be more suitable to determine the 
         stellar  behavior. 
}
 %             \label{fig:shapes}
\end{figure*}

 We emphasize, however,  that the COS dataset at hand is extremely limited, consisting of only a single in-transit observation visit and three off-transit visits on separate dates.  To definitely ascribe a stellar rotational modulation to the Si\,III 1206.5\,\AA\ and C\,II 1335\,\AA\ stellar fluxes would be premature, since there can be stochastic effects.  Furthermore, to derive a reliable transit absorption based on the observing technique used for the COS dataset, one would have to observe many times with the planet both in-transit and off-transit, to hopefully average-out potential rotational, stochastic, small flaring activity, and long-timescale magnetic  cycle variations in the stellar fluxes.  Such extensive observations are not viable with {\em HST}.  Instead, the standard transit method of directly measuring the actual stellar flux immediately before and/or after transit is by far the most direct and least ambiguous FUV transit observational method, as well as the one that would utilize the least {\em HST}  resources.Ê Already, the time variation found in the stellar fluxes in this COS dataset clearly demonstrates this argument.

\subsection{ Problems with the reported velocity distribution of the absorbers}

COS G130M transit observations of HD\,209458b have high enough sensitivity and spectral resolution to potentially determine the  velocity distribution of the absorption by the heavy species on the planet at similar S/N as in the STIS G140M  observations of the H\,I Lyman-$\alpha$ line absorption \citep{vid03,vid08,ben07,ben08,ben10}.  This has been demonstrated for both the C\,II 1335\,\AA\ doublet as well as the O\,1304\,\AA\ triplet transit observations of HD\,189733b \citep{ben13}.  Figure 2 shows the average C\,II 1335.7\,\AA\  line profiles for the four {\em HST}/COS visits of HD\,209458.  Although there may be real planetary atmospheric absorption during transit ($\phi=0$),  we find an apparent visit-to-visit variation in  the average off-transit stellar line shapes that precludes a proper evaluation of the  velocity signature in the planetary absorption.  For these reasons of potential intrinsic time variation in the stellar line shapes,  the  velocity distribution of the absorbers reported by \citet{lin10} for HD\,209458b from the COS dataset at hand is invalid.  
We note that the apparent stellar line-shape variation may be related instead to errors in the co-alignment of the spectra that was needed since the multiple grating wavelength settings were not exactly repeatable as described in Sec. 2.2, but this cannot be fully explored  with this dataset.

Proper COS transit observations are needed, with consecutive data with the planet in and out of transit, and using a single grating wavelength setting (Sec. 2.2).

%\clearpage

\section{STIS G140M and G140L Observations: No Si\,III detection}

%Table 4

\begin{table}\label{tab:obs}
\caption[]{Summary of STIS G140M observations}
\centering
\begin{tabular}{c c c c c c}
\hline\hline
 Exp. &Start  time  & Exp.     & Planet	& \\		 
 No.   & in 2001     & time     & mid.	     \\		 
 ~       & (UT)           & (sec)    & orbital phase  \\	
  \hline 							%phase
 o6e201010 & Sep 07  20:35 &      1780 &   0.980 \\
 o6e201020 & Sep 07  22:13 &       2100 & 0.000 \\
 o6e201030 & Sep 07  23:49 &       2100 & 0.019 \\
 \\
 o6e202010 & Sep 14 21:12  &     1780 &  0.973\\
o6e202020 & Sep 14 22:42  &     2100 &  0.992\\
o6e202030 & Sep 15   00:18  &     2100 & 0.011\\
\\
o6e203010 & Oct 20  02:50   &    1780 & 0.969\\
o6e203020 & Oct 20  04:20   &    2100 & 0.988\\
o6e203030 & Oct 20  05:56   &    2100 & 0.007\\
 \hline\hline 
\end{tabular}
\end{table}

%Figure 3

\begin{figure*}\label{fig:G140M}
   \centering
    \hspace*{-0.15in}
	\includegraphics[width=5.0in] {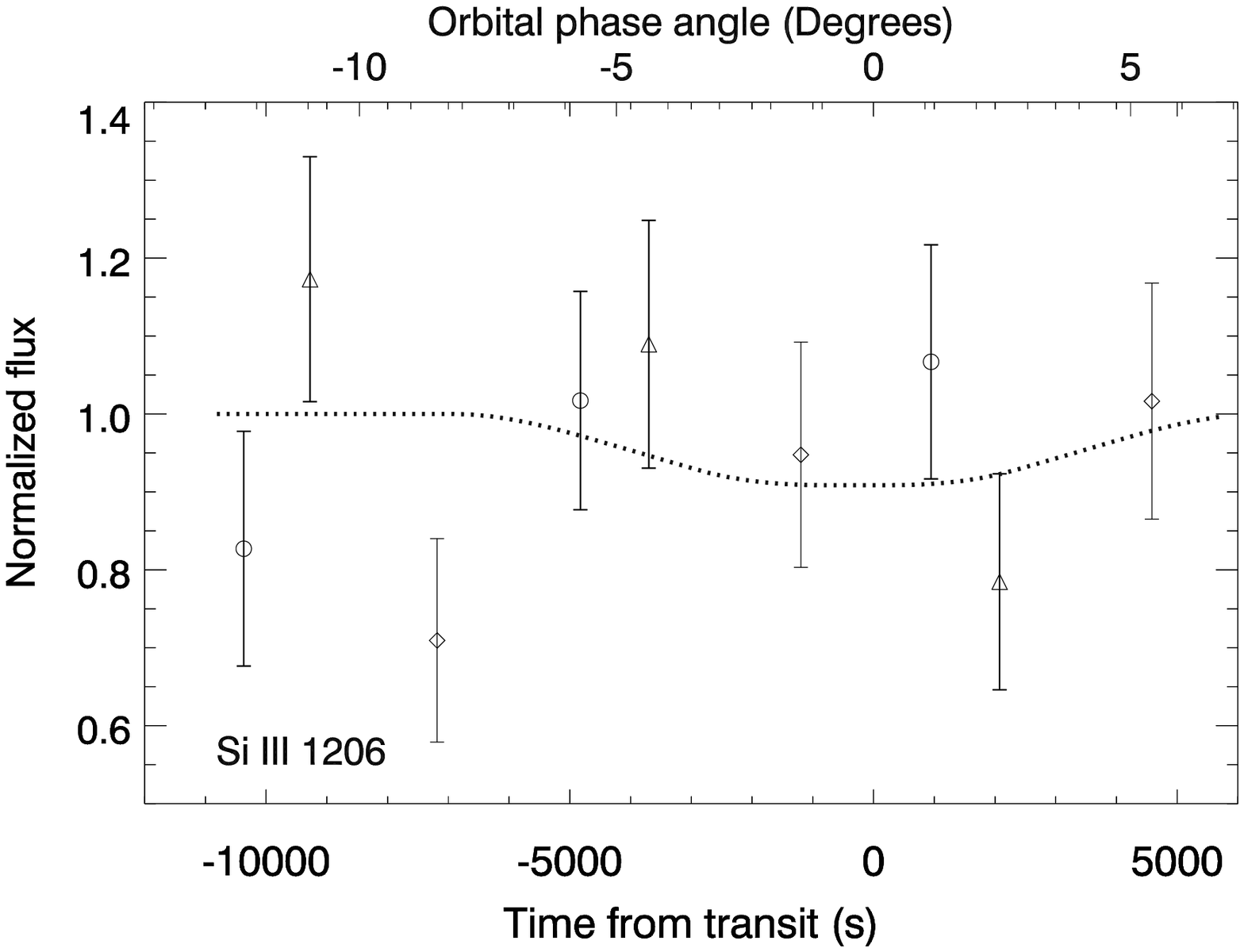} % {ballester-fig3.eps}	%
    \caption{Stellar fluxes in the Si\,III 1206.5\,\AA\   line for three transits of HD\,209458b observed with STIS G140M, the same dataset  that discovered an extended H\,I atmosphere on this planet  \citep{vid03,ben07}.    Diamonds for 7-8 Sep. 2001; triangles for 14-15 Sep. 2001; and circles for 20 Oct. 2001.  The fluxes are normalized  to the  1.56$\times10^{-15}$\,ergs\,cm$^{-2}$\,s$^{-1}$ average of the first two points. The data are integrated per  {\em HST} orbit (3 orbits per transit).  The transit light curve is drawn for an atmosphere obscuring 8.9\% of the star  \citep{ben07}.
       }
              \label{G140M}
\end{figure*}

The {\em HST} STIS medium-resolution G140M transit observations that discovered the extended hydrogen atmosphere on HD\,209458b \citep{vid03,ben07}  also sampled the Si\,III 1206.5\,\AA\ line.  With a resolving power  of $R \sim 10,000$, the line is spectrally resolved and  well separated from the wings of the  Lyman-$\alpha$ line although the S/N is relatively low.  Three transit observations were made (Table 4).   Figure 3  shows  the line fluxes per {\em HST} visit, integrated at  1206.5.07--1207.14\,\AA\ and normalized to the average of the first two points of 1.56$\times10^{-15}$\,ergs\,cm$^{-2}$\,s$^{-1}$.   No Si\,III transit absorption is  detected.  The ratio of the average of the three points within 3000\,s from mid-transit to the first  three  off-transit points yields a negative result of 1.7$\pm$18.7\% absorption.  The 1.7\% absorption, if real, would represent the 1.5\% obscuration by the  disk of the planet. 

Although the S/N is low, large temporal variations are clearly present, from visit to visit, and intra-visit. The std. dev. is 13\% for the off-transit points, and 15\% for all the data.  Thus, as with the COS observations discussed in Sec. 2.3, we find significant variation in the stellar Si\,III emission.  The  13\% std. dev.  variation of the off-transit data found with STIS  is about twice the 7.5\% value found with COS.

The original STIS G140L observations sampled four partial transits  and a negative detection in Si\,III 1206.5\,\AA\ was reported with a transit absorption of 0.0$^{+2.2}_{-0.0}$\%  by \citet{vid04}. The non-detection was later confirmed by \citet{ben10}. Although a limitation of the low-resolution G140L data is the contamination of the Si\,III line by the blue wing of the stellar Lyman-$\alpha$ line, and the S/N in the (lower sensitivity) G140M data is low, taken together, there are now seven transits observed with STIS that do not show Si\,III absorption. 

A summary of the current transit results of heavy species on HD\,209458b with STIS and COS is provided in Table 5, including the new results from STIS. We also include  transit depths from \citet{lin10} from the 2009 COS dataset  and those discussed in Sec. 2 for comparison, although all of these depths are not valid since the stellar flux around the time of  transit is unknown.   For valid COS results new   observations are required sampling the transit light curve.

%Table 5

\begin{table*}\label{tab:tr5}
\caption[]{Current transit   results for heavy species in the upper atmosphere of HD\,209458b}
\centering
\begin{tabular}{l cccc l }
\hline\hline
 Instrument   					& Species  & Species  & Species  & Species & Ref.\\
mode    						& feature (\AA) & feature (\AA)  & feature (\AA)  & feature (\AA)  & \\  
 \hline \hline
 \multicolumn{6}{l}	 {\em  Absorption depths from transit light curve FUV observations}\\
%\hline
{ } & {O\,I 1304}   & {C\,II 1335}  & {Si\,III 1206.5} 	 & {Si\,IV 1393.7} &  \\
{ } &  triplet  &   doublet &   line	 &    line \\
{STIS G140L} & 12.8$\pm$4.5\%	  &	7.5$\pm$3.5\%	  &	 0.0$^{+2.2}_{-0.0}$\%  &	 0.0$^{+6.5}_{-0.0}$\% &Vidal-Madjar et al. (2004)$^a$\\
{STIS G140L} &  10.5$\pm$4.4\%	  &	7.4$\pm$4.7\%	  &	confirmed	  &	confirmed	  & Ben-Jaffel \& Hosseini (2010)$^a$\\
{STIS G140M}& n/a	  &	n/a  &	1.7$\pm$18.7\%  	  &	n/a	  &this work$^a$\\ 
\hline
 \multicolumn{6}{l}	 {\em  Invalid  depth estimates from COS 2009 dataset from in-transit data against data from other dates}\\
{ } 				&	& {C\,II 1335}  		& {Si\,III 1206.5}		& {Si\,IV 1393.7} 		&  \\
{ } 				&      &   doublet 			&   line	 			&    line \\
{COS G130M}		&  	&	7.8$\pm$1.3\%	&	8.2$\pm$1.4\%	&	0.00$\pm$0.01\%	&  Linsky et al. (2010)$^b$\\ 
{~$''$, this work} 	&  	&	9.3$\pm$2.4\%	&	8.6$\pm$2.7\%	&	3.1$\pm$4.2\%	& vs off-tr. ave, photon noise$^c$\\ 
{~$''$, this work} 	&  	&	5.3$\pm$2.9\%	&	-1.9$\pm$3.5\%	&	1.5$\pm$4.9\%	& vs $\phi=0.27$, photon noise$^d$\\ 
 {~$''$, this work}	&	&	9.3$\pm$8.4\%	&	8.6$\pm$9.5\%	&	3.1$\pm$12.1\%	& vs off-tr. ave, w intra-visit var.$^e$\\ 
{~$''$, this work} 	&  	&	5.3$\pm$9.8\%	&	-1.9$\pm$12.0\%	&	1.5$\pm$16.8\%	& vs $\phi=0.27$, w intra-visit var.$^e$\\ 
\hline\hline
\end{tabular}
\begin{list}{}{}
 \item[$^{a}$] The STIS G140L and G140M results were respectively derived from 4 and 3  transit observations.
 \item[$^{b}$] Invalid  "transit depth" reported by 
 \citet{lin10},
 %Linsky et al. (2010), 
 not from a transit light curve but from the average of data with the planet off-transit at orbital phases $\phi=0.27$, 0.72 and 0.49, and with errors for photon noise only.
 \item[$^{c}$]As in item {\em b}, for comparison with the Linsky et al.  results, with photon  noise only. 
\item[$^{d}$] Sample invalid "transit depth" from contrasting the in-transit data against  off-transit observation at $\phi=0.27$.  A negative detection is obtained in Si\,III, and a smaller transit depth is obtained in C\,II.  
\item[$^{e}$] Invalid "transit depths" derived as in items {\em c} or {\em d} but  including  the intra-visit standard-deviation variations (Table 2) in the errors.  No previous HD\,209458b UV transit study has included this variation.  Full transit light curves are needed for a proper assessment of these variations and thus of the true potential of the COS transit measurements of HD\,209458b.
\end{list}
\end{table*}

\section{Discussion: New light on contribution of UV studies to exoplanetary atmospheric sciences}

The abundance of heavy species in the upper atmosphere has major implications on all pressure levels from the top to the bottom of a  planetary atmosphere.  The link of the FUV transit observations with properties of the planet and its lower atmosphere are often underplayed. The mere detection of extended O\,I and C\,II absorption in HD\,209458b's upper atmosphere with STIS G140L (at low spectral resolution) already requires solar or supersolar abundances, a firm result that is in sharp disagreement with the recent inference of potential $\times$20--125 low metallicity on this planet \citep{mad14}.  This low metallicity has been raised as one of two explanations for an inferred  low H$_2$O  abundance in the planet's lower atmosphere: either a low metallicity and C/O $<$ 1 since water would be the major oxygen-bearing species in this case, or; a high metallicity but C/O $>$ 1.  A $\times$20--125 low metallicity would be highly unexpected on a hot Jupiter. 
%\citep[Note that HD\,209458  has a solar metallicity {[Fe/H]}=--0.0$\pm$0.2;][]{maz00}
(Note that HD\,209458  has a solar metallicity [Fe/H]=--0.0$\pm$0.2; Mazeh et al. 2000.)   
The goal of  \citet{mad14} is to explain a low contrast 1.4\,$\mu$m band absorption reported for this planet from {\em HST} Wide-Field Camera 3 transit data, yet, it is possible that some level of haze scattering is present \citep{dem13},  as well as that the data need further work (T. Evans et al., in progress).  Since  species in the upper atmosphere  derive from dissociation products in the lower atmosphere, issues of metallicity and C/O ratio are directly testable with FUV observations -- provided high quality  data resolving the transit absorption line shapes are obtained.

%related to the planet formation scenarios 

The presence of Si\,III in the upper atmosphere of HD\,209458b would have implied that silicon gas is not effectively depleted by cloud condensation in the lower atmosphere \citep{kos13a,kos13b,lav14}.  This turns out to possibly still be the case as demonstrated by the recent detection of $6.2\pm2.9$\% absorption (or $8.8\pm2.1$\% if there was still post-egress absorption)  by extended Mg\,I on this planet at near-solar abundance \citep{vid03}, because  magnesium and silicon  condense together into  silicate grains  of fosterite and enstatite  \citep{vis10}. The detection of  magnesium in the thermosphere of HD\,209458b indicates that the balance of day-to-night temperature  \citep{sho09,mos11,cro12} against the strong vertical and  horizontal 3D dynamics \citep{sho09,spi09,par13} impedes significant condensation and settling of refractory silicate species on this planet \citep{kos13a,kos13b,lav14}.  If condensation takes place, such as on the nightside, the  transport of fine silicate grains onto  hotter high-altitude dayside regions  may allow  for effective sublimation on this hot Jupiter.  Such a strong  transport of fine grains  has been found to be  possible on HD\,209458b   based on  3D global circulation modeling that included test particles and found a strong circulation with large effective eddy coefficients of for example $K_{zz}\sim10^{10}$\,cm$^2$s$^{-1}$ at $P\sim1$\,mbar \citep{par13}. 
 
With respect to the FUV and NUV observations and the upper-atmosphere modeling, the disagreement lies in the  predicted Si\,III and observed Mg\,I ionization states, such that silicon might be present mainly as either  Si\,I  or Si\,II.  In the latter case charge-exchange reaction rates invoked to convert part of the Si\,II into Si\,III  would need to have been  overestimated so that  no significant Si\,III 1206.5\,\AA\ absorption  is detected in transit.  The charge exchange of Si\,II with protons ($ {\rm Si}^{+} + {\rm p}^+ \rightarrow {\rm Si}^{++} + {\rm H}$) was first proposed by \citet{lin10} as the main source  of Si\,III since this reaction has been found to be relevant for the Sun.  In their detailed upper-atmosphere models for HD\,209458b that  included heavy metal species, \citet{kos13a,kos13b} confirmed that this reaction is key in producing detectable amounts of Si\,III to fit the $\sim$8\% transit depths reported by Linsky et al. from the COS dataset.  
However, our new STIS G140M negative finding  for Si\,III on this planet that is consistent with the previous negative detection by \citet{vid04} (Table 5),  poses new questions into the modeling and/or the state of the upper atmosphere of this planet.  It may be that the model temperatures are too high since temperatures above $\sim$15,000\,K  are required for the Sun  to produce significant Si\,III abundances \citep{bal80} and these temperatures do not necessarily apply to this planet.  The model temperature depends on the chemical composition, on the stellar XUV input and the heating efficiency,  and on the velocity of the outflow since it can  adiabatically cool the atmosphere, although there are always uncertainties in these calculations.  We find in the literature that only models with high stellar XUV input and high heating efficiency seem  to reach high temperatures of 15,000\,K and beyond,  such as  all the model cases in \citet{tia05}, the sub-sonic cases in \citet{gar07}, and the models in \citet{kos13a,kos13b} that explicitly assume a $\times$10 or $\times$100 higher solar XUV input for the star.  Other models reach about 12,000--14,000\,K peak temperatures, such as the \citet{yel04} model, the super-sonic cases in   \citet{gar07},  and  the \citet{kos13a,kos13b} models with solar XUV input.  
In the modeling of the upper atmosphere of a non-magnetized HD\,209458b and its interaction with the stellar wind,  \citet{mur09} find that the peak temperature reaches $\sim$10,000\,K for a solar XUV flux,  while for a 1000 times larger XUV flux typical of young T-Tauri stars the peak temperature reaches only slightly larger values due to radiative H\,I Lyman-$\alpha$ cooling rather than by the expansion of the gas.
In a separate work we are  finding evidence  that the star HD\,209458 has a surprising  long-term low activity, and this indicates that the star provides  a significantly lower XUV input to the planet than assumed in all previous modeling \citep{bal15}.  In  that work we apply the detailed photochemical hot-Jupiter upper-atmosphere models of   \citet{gar07} that include detailed photochemistry for both ion and neutral species of H, D, He, O, C and N and for which we have updated  our best current estimate of the stellar X-ray and UV input.   The modeling  finds indeed peak atmospheric temperatures (for solar abundances) of only $\sim$10,000\,K.  Given the Mg\,I detection, and accompanying non-detection of Mg\,II \citep{vid13},  the dominant silicon ionization state  may be Si\,I since Mg\,I and Si\,I have similar ionization potentials.  Nevertheless, the finding that Mg\,I dominates in the upper atmosphere of this planet  is highly unexpected, since significant photoionization ($\lambda<1621$\,\AA) should be present while  Mg\,II was not positively detected. This  seems to require very large electron densities for the dielectric recombination required to counteract photo-ionization:  
for example, at the reference altitude of  $r=3\,$R$_{p}$  \citet{bou14} require  at least two orders of magnitude higher electron densities than estimated from  current  hydrodynamic models \citep{gar07,kos13a,kos13b,guo13,lav14,bal15}.  This is something that is rather hard to explain.   

Another finding is that the Mg\,I 2853.0\,\AA\ line absorption seems to be blue-shifted to --62 to --19\,km\,s$^{-1}$ \citep{vid13}. This finding is quite compelling to some and unexpected to others, as per the discussion in the Introduction related to the H\,I Lyman-$\alpha$ line.   
Again, the large electron densities required for dielectronic recombination  and the fast outflows needed at the exobase to explain the Mg\,I blue-shifted absorption by stellar radiation pressure \citep{bou14} are difficult to  reconcile with current upper atmosphere models. Yet,  if confirmed,  the Mg\,I blue-shifted absorption reveals a net anti-solar motion that may apply to one or more components of the upper atmosphere.   Do the O\,I and C\,II species also show a net blue-shifted absorption?  Are the motions of the ions and neutrals fully coupled  \citep[e.g.,][]{gar07,tra11,ada11} or are they decoupled at large enough distances and low collision rates \citep[e.g.,][]{kos14}? The latter could be expected from effects by different radiation pressure, charge exchange with the stellar wind protons, close and open planetary magnetic-field lines and  decaying field strength with distance, and net magnetospheric currents and convection.  Velocity-resolved transit data on the C\,II 1335 and O\,I 1304\,\AA\ lines, even if only separating the blue- and red-component  absorptions, would provide much needed tools for characterizing  the state of the upper atmosphere of HD\,209458b that is still an enigma.  Such data can be obtained with COS.  

\section{Conclusions}

We have presented new STIS G140M transit data reinforces the original results with STIS G140L that Si\,III ions are not detected in the thermosphere of HD\,209458b.  Silicon may still be present in the upper atmosphere, but at a lower ionization state that may even be Si\,I, based on the independent {\em HST} detection of extended Mg\,I on this planet which also indicates that silicate condensation and cold trapping is not 100\% effective in the lower atmosphere.   
The models for HD\,209458b by \citet{kos13a,kos13b} that include refractory species already predict Si\,II to be dominant silicon species in the upper atmosphere, so what is needed as a first step is to limit the yield for Si\,III.  A second step much harder to address would be whether the silicon is in the Si\,II or Si\,I state, given the positive Mg\,I but negative Mg\,II detection on HD\,209458b by \citet{vid13} and that Si\,I and Mg\,I have similar ionization potentials. The lack of  a positive  Si\,III detection in the extended atmosphere of the planet reported in this work, together with the positive Mg\,I detection and the long-term low  activity  now being identified for the star \citet{bal15} indicate that a revision to our present understanding of the upper atmosphere of this planet is needed.  
Much new insight can be obtained with new COS FUV transit observations.

The existing COS G130M observations of the HD\,209458 system were not standard transit observations and significant stellar flux  variations found in the data invalidate previously reported transit results. Proper FUV transit observations with COS are needed, sampling the transit light curve (at a single grating wavelength setting).  These observations will  independently confirm and measure transit depths in the O\,I and C\,II species at high S/N, and furthermore resolve the velocity distribution of the O\,I and C\,II species with major implications on upper atmospheric dynamics, energetics and magnetospheric processes, as well as on the  species abundances that are also relevant for the characterization of the planet and its lower atmosphere.  In this frame, the detection of OI and C II in the upper atmosphere of HD\,209458b using only unresolved lines already requires solar or supersolar abundances that are in  disagreement with  a very low metallicity as recently considered in the literature.  In the near future, upper atmospheric UV studies of exoplanets may prove key for unraveling the so far elusive properties of low mass exoplanets.   

\acknowledgments
We are grateful to Brian York and Justin Ely at the HST Help Desk  for input on COS observations and data re-processing.  
We thank F. Pont, D. Sing, P. Lavvas, A. Garc\'ia Mu\~noz, A. Showman, V. Parmentier, and T. Koskinen for interesting conversations on metallicity, condensation, circulation, and upper atmospheres of hot Jupiters applicable to this work.   We also thank  the referee for helpful suggestions on the manuscript.
LBJ acknowledges support from CNES, Universit\'e Pierre et Marie Curie  and the Centre National de la Recherche Scientifique in France.  He also acknowledges his current appointment as a Visiting Research Scientist at the Dept. of Planetary Sciences and the Lunar and Planetary Laboratory of the University of Arizona.  GEB  was partially funded by the Space Telescope Science Institute under grants HST-GO-12473.01-A and HST-AR-11303.01-A  to the University of Arizona. 

{\it Facilities:} \facility{HST (STIS)}, \facility{HST (COS)}.

\clearpage

\end{document}